\documentclass[10pt, conference, letterpaper]{IEEEtran}
\usepackage{float}
\usepackage{amsmath}
\usepackage{amssymb}
\usepackage{amsthm}
\usepackage{tabularx}
\usepackage{graphicx}
\usepackage{subfigure}
\usepackage{authblk}
\usepackage[linesnumbered, ruled]{algorithm2e}

\usepackage{threeparttable}
\usepackage{slashbox}
\usepackage{algorithmic}
\usepackage{tabularx}
\usepackage{booktabs}
\usepackage{amsfonts}
\usepackage{dsfont}
\usepackage[all]{xy}

\def\squarebox#1{\hbox to #1{\hfill\vbox to #1{\vfill}}}

\newtheorem{thm}{Theorem}
\newtheorem{cor}[thm]{Corollary}

\newtheorem{defn}{Definition}

\newtheorem{obsv}{Observation}
\newtheorem{exam}{Example}

\begin{document}

\title{Online Query Scheduling on Source Permutation for Big Data Integration}

\author{Zimu Yuan, Shusheng Guo \\ 
University of Chinese Academy of Sciences, China\\
Institute of Computing Technology, Chinese Academy of Sciences, China\\
\{yuanzimu, guoshusheng\}@ict.ac.cn
}

\maketitle \thispagestyle{empty}

\begin{abstract}
Big data integration could involve a large number of sources with unpredictable redundancy information between them. The approach of building a central warehousing to integrate big data from all sources then becomes infeasible because of so large number of sources and continuous updates happening. A practical approach is to apply online query scheduling that inquires data from sources at runtime upon receiving a query. In this paper, we address the Time-Cost Minimization Problem for online query scheduling, and tackle the challenges of source permutation and statistics estimation to minimize the time cost of retrieving answers for the real-time receiving query. We propose the online scheduling strategy that enables the improvement of statistics, the construction of source permutation and the execution of query working in parallel. Experimental results show high efficiency and scalability of our scheduling strategy.
\end{abstract}




\section{Introduction}
Big data not only means a large volume of data, but also indicates variety that data can be from a large number of sources. The integration of big data that unifies varieties of data from widely-distributed sources can act as the foundation of information conformity for applications. For this big data integration, it has two major changes compared with traditional data integration. First, continuous updates may exist in a fair amount of data sources. As a consequence, the traditional way of building a central warehousing to integrate data from sources would become infeasible. A practical way is to apply online scheduling that inquires data from sources at runtime upon receiving a query. Second, a large proportion of unpredictable redundant data exist between sources. Thus, the query results returned from different sources should be judged, removing the repetitive ones. To deal with these two challenges, an intelligent online query scheduling could be designed to answer the query with non-redundant results in the least possible time.

\begin{figure}
\centering
\includegraphics[width=3.4in]{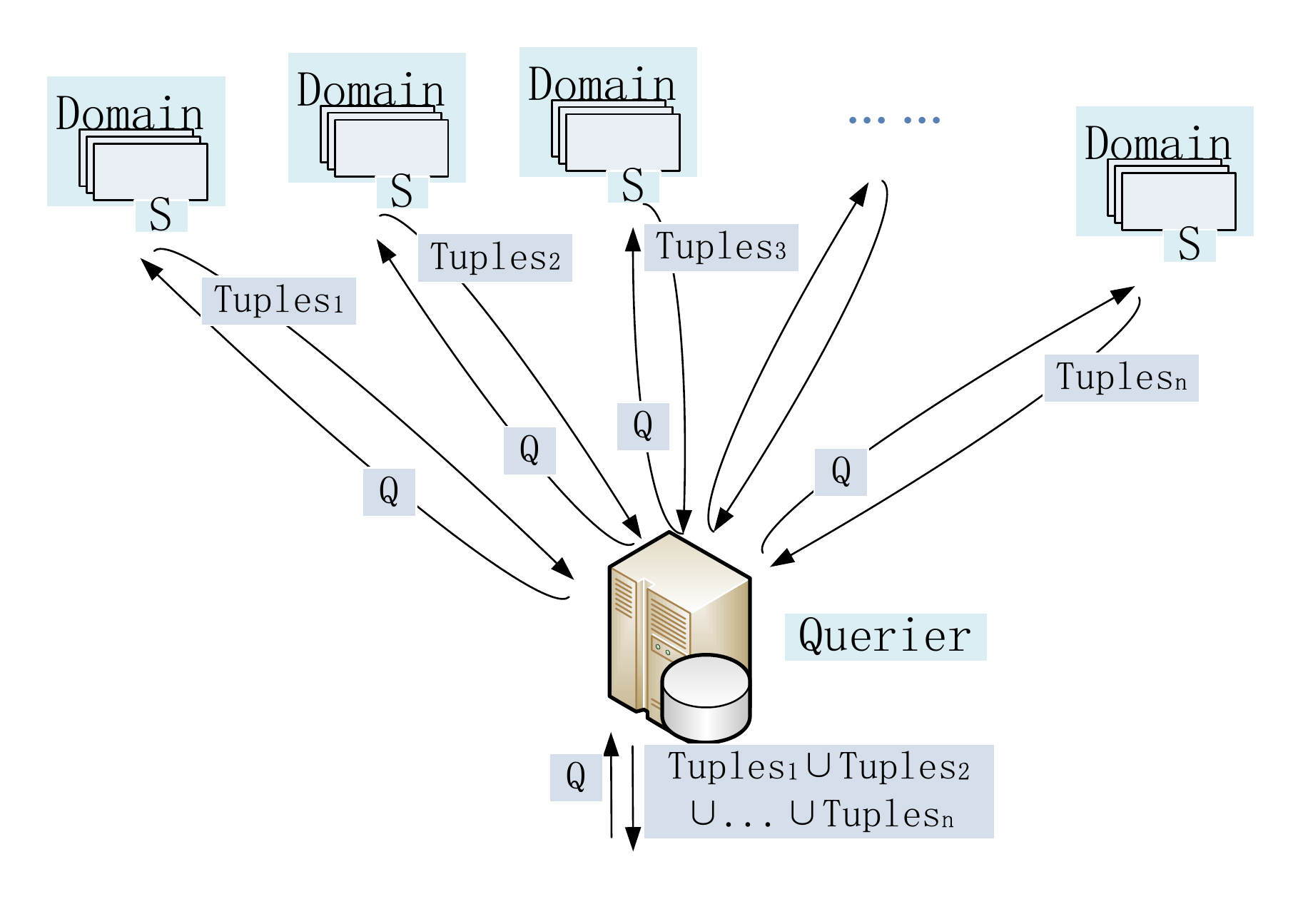}\\
\caption{\textrm{The scenario of online query scheduling for integration}} \label{newfig}
\end{figure}

In this paper, we set out to address this online query scheduling problem. In its scenario, as illustrated in Figure \ref{newfig}, the querier that does not cache any data itself and the domains that act as data sources are independent each other. For query scheduling, the querier may firstly arrange the domains in a permutation. Then, upon receiving a query, the querier sequentially inquires each domain (or few domains in parallel) at a time following the permutation. When obtaining new results from a domain, the querier compares them with previous results received from other domains, and removes the repetitive ones. Finally, the querier returns the results after receiving enough ones. We also found that the scenario depicted in Figure \ref{newfig} is common in research and practical application. In application, this style of online query on multiple sources is usually seen in Aggregator. For example, Google News, a news aggregator, watches updates from more than 4500 worldwide news sources, and exhibits non-redundant news to the readers; \emph{KAYAK.com}, a trip aggregator, shows similar trips obtained from hundreds of travel sites. More aggregator applications can be referred in \cite{aggapp}. Besides, in domain-centric research, related studies mainly focus on integration of the knowledge about a topic from widespread sources. As shown in the experiment of \cite{Dalvi2012}, about 5000 sources are needed to be queried on for acquiring 95\% knowledge of a topic due to a large portion of redundant knowledge. Also, similar results in other topics are given in \cite{Salloum2013}.

For reduction of query time, in our experience of online scheduling, we have found that the critical point lies in the permutation of sources. Each data source is with a different access time, a different transfer time, and a different proportion of intersection data (namely redundant information) with other sources. Choosing a good permutation of sources can reduce the total time cost for a query, especially with much more significantly reduction for the query in a long run on a large number of sources. To better describe the permutation choice, we give a simple example on query $Q_k$ (namely a query $Q$ for acquiring $k$ tuples)
here:

\begin{figure*}[!htb]
\begin{minipage}[]{0.28\textwidth}
\centering\includegraphics[width=1.6in]{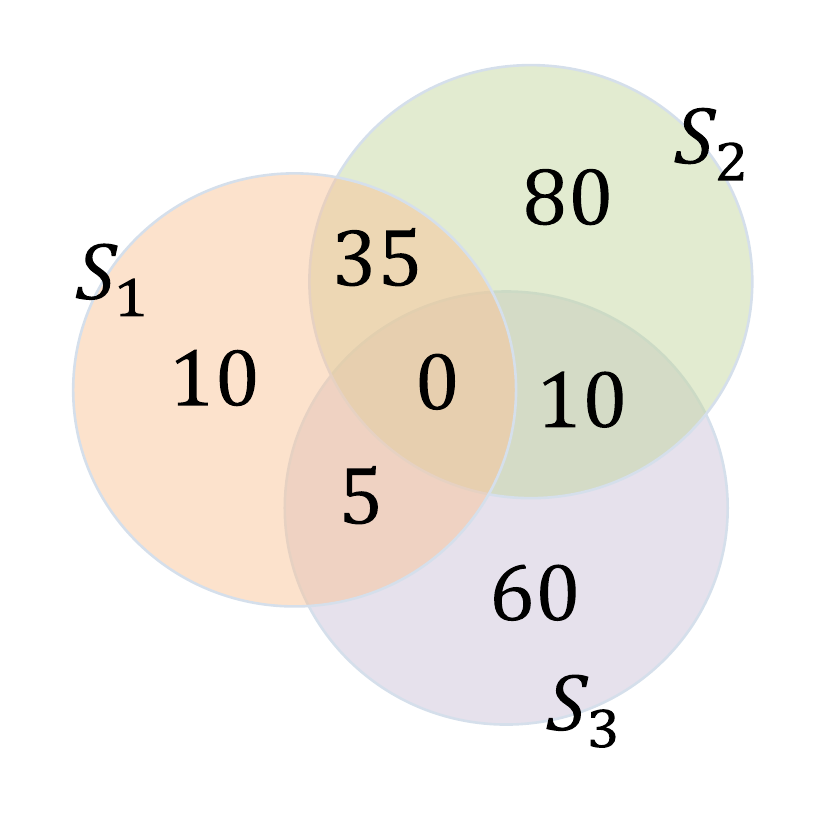}
\caption{\textrm{Venn diagram of $S_1$, $S_2$ and $S_3$}}\label{f1}
\end{minipage}
\begin{minipage}[]{0.35\textwidth}
\centering\includegraphics[height=1.65in,width=2.4in]{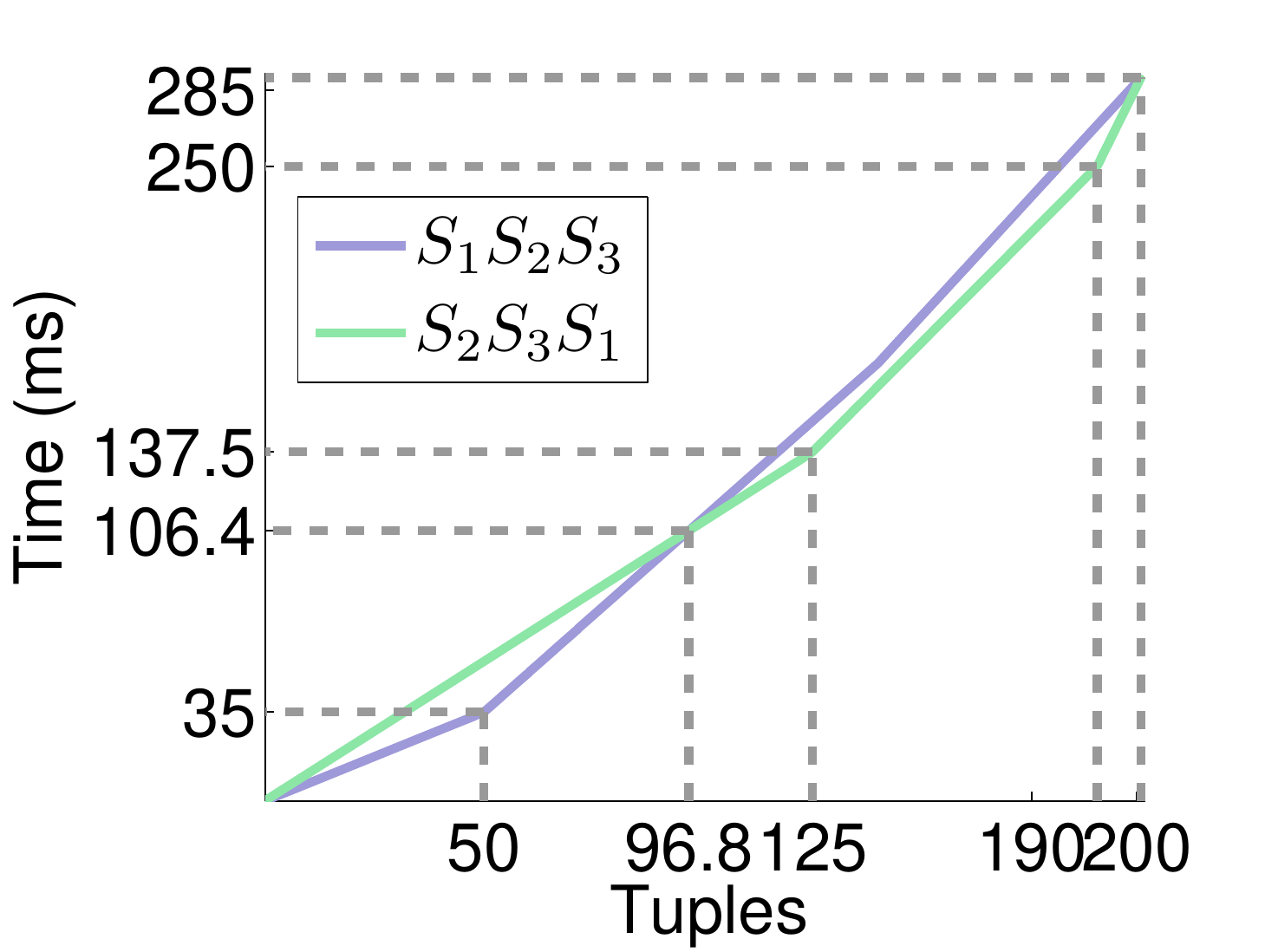}
\caption{\textrm{Time cost of the permutations}}\label{f2}
\end{minipage}
\begin{minipage}[]{0.36\textwidth}
\centering\includegraphics[width=2.55in]{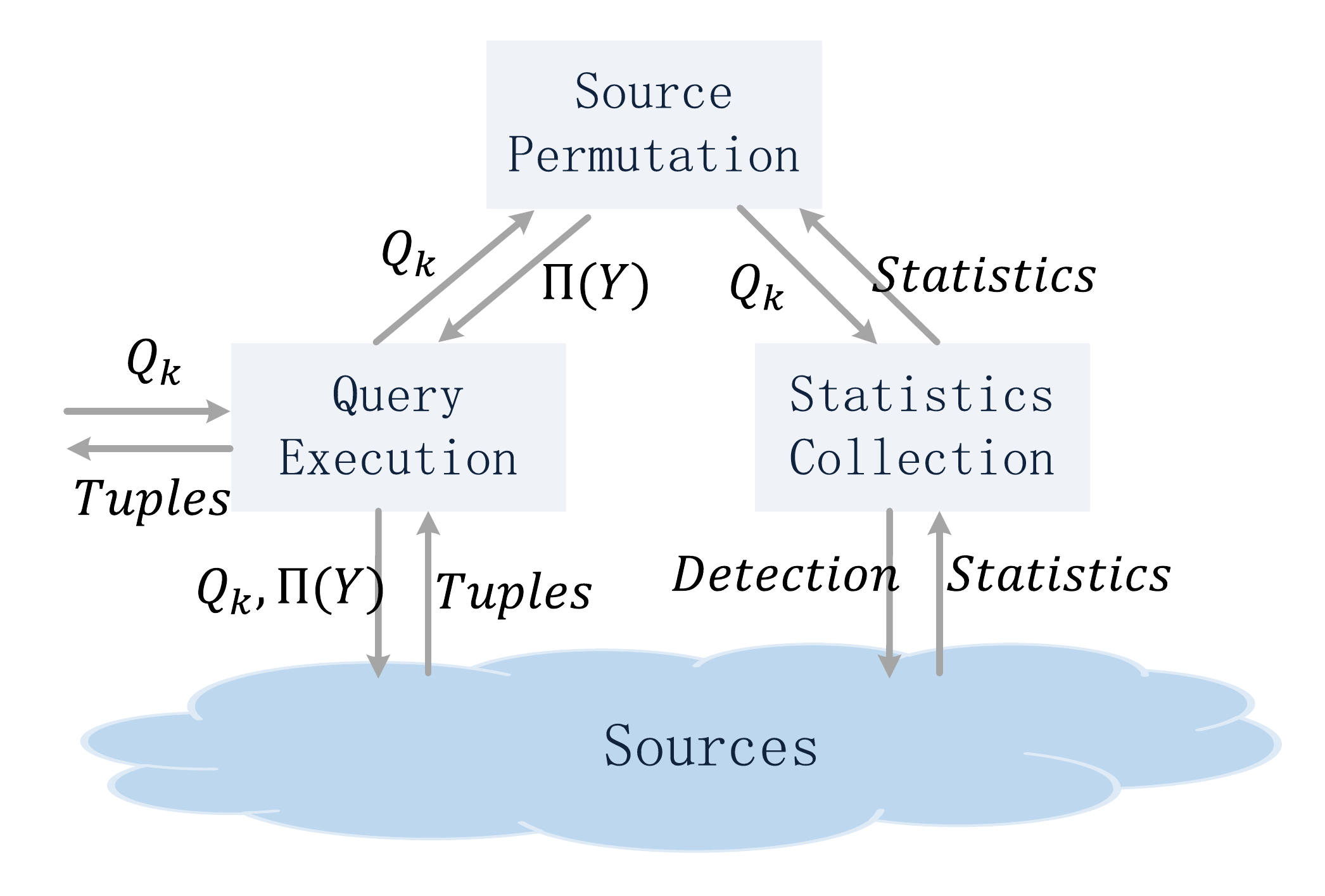} 
\caption{\textrm{The architecture of online query scheduling}}\label{f3}
\end{minipage}
\end{figure*}

\begin{exam}
Consider three sources $S_1$, $S_2$ and $S_3$ shown in Figure \ref{f1}. For simplicity, set the access time $ta_1=ta_2=ta_3=0$ and the transfer time of retrieving a tuple $tr_1=0.7 \; ms$, $tr_2 = 1.1 \; ms$ and $tr_3 = 1.5 \; ms$ for $S_1$, $S_2$ and $S_3$ respectively. Upon receiving a query $Q_k$, the querier starts to estimate the statistics of sources, and has the estimated value of statistics that $S_1$, $S_2$ and $S_3$ have $|S_1|=50$, $|S_2|=125$ and $|S_3|=75$ result tuples for $Q$ respectively, and the intersection tuples $|S_1 \cap S_2|=35$, $|S_1 \cap S_3|=5$, $|S_2 \cap S_3|=10$ and $|S_1 \cap S_2 \cap S_3|=0$. Then, the querier has the following optimal permutations $\Pi_{opt}$ on time cost (permutation $S_1$ and $S_1S_2$ can be seen in curve $S_1S_2S_3$, and permutation $S_2$, $S_2S_3$ and $S_2S_3S_1$ can be seen in curve $S_2S_3S_1$ in Figure \ref{f2}):
\begin{equation} \label{e_example}
\Pi_{opt} =
\begin{cases}
\displaystyle S_1 &  0<k \leq 50 \\
\displaystyle S_1S_2 & 50<k \leq 96 \\
\displaystyle S_2 & 96<k \leq 125 \\
\displaystyle S_2S_3 & 125<k \leq 190 \\
\displaystyle S_2S_3S_1 & 190<k \leq 200 \\
\end{cases}
\end{equation}
For instance, suppose that the querier receives a query $Q$ of $125$ tuples and only starts one thread to process it, the optimal permutation $S_2$ has the query time cost $tr_2|S_2|=137.5 \; ms$. In contrast, given another permutation $S_1S_2$, only $50$ tuples are obtained after querying on $S_1$; then, the querier start to inquire $S_2$ for another $125-|S_1|=75$ tuples, and the expected time of receiving a non-repetitive tuple from $S_2$ is $\frac{tr_2|S_2|}{|S_2|-|S_1 \cap S_2|} \approx 1.5 \; ms$; finally, following permutation $S_1S_2$, the querier returns the result of $125$ tuples with time cost $tr_1|S_1|+1.5(125-|S_1|) \approx 147.5 \; ms$, greater than the optimal time cost $137.5 \; ms$.
\end{exam}

We named the problem addressed as Time-Cost Minimization Problem (TMP, formally defined in Section \ref{model}). In implementation, TMP has two major difficulties: statistics estimation (about the intersection between sources) and optimal permutation based on statistics. Concerning on these two difficulties, we make the following contributions\footnote{For quickly understanding our paper, please refer to Section \ref{architecture}: Architecture (Figure \ref{f3}) and Section \ref{strategy}: Online Query Strategy (Figure \ref{fig5}).}:
\begin{itemize}
  \item We prove that TMP is NP-complete, and propose OnlinePerm algorithm that constructs the permutation $\Pi(Y)$ with the time cost $T(Q_k(\Pi(Y)))$ $\frac{k \sum_{i=1}^{l} (ta_{i}+tr_{i}|S_{i}|)}{|Y| \sum_{i=1}^{u(i_k)} (ta_{u(i)}+tr_{u(i)}|S_{u(i)}|)}$-approximately to the optimal time cost $T(Q_k(\Pi_{opt}(Y)))$ for a query $Q$ of retrieving $k$ tuples.
  \item We present the mechanism of two-stage detection for statistics collection. Especially, to avoid the exponentially growing complexity of detection when the number of sources increases, the statistics collection mechanism apply the pruning techniques that only probes the critical statistics on-the-fly.
  \item We propose the online scheduling strategy that enables the improvement of statistics collection, the permutation construction of OnlinePerm algorithm and the execution of query on sources working in parallel, so as to reduce the total time cost for the query.
  \item We conduct experiments on physical sources. The experiment results show that our scheduling strategy is scalable and can significantly reduce the time cost.
\end{itemize}

The rest of this paper is organized as follows. Section \ref{related} presents the related work. Section \ref{overview} gives the overview of online query scheduling. Section \ref{sp} presents the OnlinePerm algorithm for source permutation. Section \ref{sc} proposes the mechanism of two-stage detection for statistics collection. Section \ref{strategy} describes of the scheduling strategy. Section \ref{experiment} presents the experiment results. Section \ref{conclusion} concludes this paper.

\section{Related Work} \label{related}
There exist some research work \cite{Bleiholder2006}\cite{Dong2013}\cite{Florescu1997}\cite{Nie2005}\cite{Salloum2013}\cite{Sarma2011}\cite{Trushkowsky2013}\cite{Vassalos1998} on consideration of intersection or redundancy information for source selection. In \cite{Florescu1997}, sources were classified into domains (e.g. Journal paper, Conference paper), and probabilistic intersections between domains were considered to select sources with more answers. In \cite{Vassalos1998}, the challenges on calculating the exponential number of source intersections were discussed to provide useful knowledge for source permutation. In \cite{Nie2005}, the StatMiner system was proposed to learn the union and intersection statistics between classes (including sources and quieres) and provide static selection of sources with more relevant answers for each class. In \cite{Bleiholder2006}, the maximize number of duplicates were found between sources to answer the Join queries for entities. In \cite{Sarma2011}, the copying relationships between sources known in advance were considered to determine the permutation of sources with the target of cost-minimization and maximum-coverage. In \cite{Dong2013}, possible intersections or duplicates between sources are studied to estimate the expected value for data items, and thus data integration system could construct a permutation of sources by the accuracy of data items. In \cite{Trushkowsky2013}, the answers for a query were estimated by the statistics (including intersection knowledge) obtained by other queries for the crowdsourced system, and thus a permutation of sources can be given by sorting the sources with the number of answer tuples in descending order. In \cite{Salloum2013}, the OASIS system was proposed to online collect intersection statistics and find a permutation of sources to maximize the Area-under-the-curve value for retrieving all answer tuples from sources.

\textbf{Summary of differences}: Our work differentiates with related literatures in following aspects: (1) we take the real-time receiving query $Q$ of $k$ tuples as input, (2) we focus on the Time-Cost Minimization problem for source permutation, (3) we propose a two-stage detection mechanism that combines the offline and online collection processes to estimate intersection statistics between sources, and (4) we present the online scheduling strategy that enables source permutation, statistics collection and query execution working in parallel.

\section{Overview} \label{overview}

\subsection{Model and Problem} \label{model}
Given a set of sources $Y=\{S_1,S_2,...,S_l\}$ and a permutation $\Pi(Y)$ for a query $Q$, the query rate $v_i$ of a source $S_i \in Y$ can be written as follow:
\begin{equation} \label{e_v}
v_i=\frac{|S_i|-|\cap S_i|}{ta_i+tr_i|S_i|}
\end{equation}
In (\ref{e_v}), $|S_i|$ is the total number of result tuples for $Q$ in source $S_i$; $|\cap S_i|$ is the intersection tuples that has been transferred to the querier from other sources (that is, these intersection tuples are in $|S_j \cap S_i|$, where $S_j$ is prior to $S_i$ in $\Pi(Y)$); $ta_i$ and $tr_i$ denote the access time and the transfer time of a tuple from $S_i$ to the querier respectively.

Suppose that the permutation $\Pi(Y)$ has $\Pi(Y)=S_{(1)}S_{(2)}...S_{(i_k)}$, $ \sum_{i=1}^{i_k-1} (|S_i|-|\cap S_i|) < k \leq \sum_{i=1}^{i_k} (|S_i|-|\cap S_i|)$, and only one query thread is running in the querier, we have the average rate $v_{avg}= \frac{\sum_{i=1}^{i_k} (|S_{(i)}|-|\cap S_{(i)}|)}{\sum_{i=1}^{i_k} (ta_{(i)}+tr_{(i)}|S_{(i)}|)}$ and the time cost $T(Q_k(\Pi(Y)))=\frac{k}{v_{avg}}$. The optimal permutation $\Pi_{opt}(Y)$ has the least time cost compared to any other permutations. We define the Time-Cost Minimization Problem as follow:

\begin{defn}[Time-Cost Minimization Problem (TMP)]
Given a query $Q$ of $k$ tuples and a set $Y=\{S_1,S_2,...,S_l\}$ of sources, find the optimal permutation $\Pi_{opt}(Y)$ of sources, having time cost $T(Q_k(\Pi_{opt}(Y))) \leq T(Q_k(\Pi(Y)))$ for any other $\Pi(Y)$.
\end{defn}

\subsection{Architecture} \label{architecture}
The architecture of online query scheduling (Figure \ref{f3}) mainly contains three components:
\begin{itemize}
  \item Source Permutation (SP): Upon receiving a query $Q_k$, SP repeatedly improves the permutation $\Pi(Y)$ for $Q_k$ based on the continuously collected statistics provided by the Statistics Collection (SC) component. The permutation $\Pi(Y)=S_{(1)}S_{(2)}...S_{(i_k)}$, $ \sum_{i=1}^{i_k-1} (|S_i|-|\cap S_i|) < k \leq \sum_{i=1}^{i_k} (|S_i|-|\cap S_i|)$, contains sources having a total of no less than $k$ tuples for $Q_k$, and the set $Y_e$ contains the remaining unselected sources.
  \item Statistics Collection (SC): SC collects the statistics from all sources. Its process of collection is divided into two stages. In the Initial-detection stage, SC generates a query $\widehat{Q}$ that retrieves all tuples from all sources (or samples from all sources) in $Y$ to obtain a general statistics. Then, the statistics of any other query can be regarded as a subset of the statistics of $\widehat{Q}$. When receiving the query $Q_k$, SC starts the Online-detection stage. In this stage, the statistics of $Q_k$ are firstly estimated by the statistics of $\widehat{Q}$, and then continuously improved by the online detection results on sources.
  \item Query Execution (QE): QE runs the query threads to retrieve the tuples of results for $Q_k$ from sources following the permutation $\Pi(Y)$ constructed by SP. When the querier has already received $k$ tuples for $Q_k$ (or all sources have been queried), QE sends a signal to terminate the running process in SP and SC.
\end{itemize}
Next, we present SP and SC in Section \ref{sp} and \ref{sc} respectively. Then, in Section \ref{strategy}, We describe the scheduling strategy that enables SP, SC and QE working in parallel.

\section{Source Permutation for Minimal Time Cost} \label{sp}
In this section, we firstly prove that the TMP is NP-complete, and then provide two observations that reveal the correlation between the query rate and the residual tuples. Finally, we propose the OnlinePerm algorithm for source permutation based on these two observations.

\begin{thm}
TMP is NP-complete.
\end{thm}
\begin{IEEEproof}
We prove the NP-hardness of TMP with the Set Cover Problem. Given a universal set and a set of subsets whose union equals the universal set, the Set Cover Problem is to find the smallest number of $m$ subsets whose union equals the universal set. The Set Cover Problem can be instantiated as follows. Suppose that the universal set $Y^{'}$ contain $|Y|+|Y|^2$ elements, $|Y|$ of which are the union of tuples for query $Q$ from $S_i \in Y$, $i=1,2,...,l$, and the other $|Y|^2$ of which are from the newly generated tuples for query $Q$. Create $l$ subsets $S^{'}_1$,$S^{'}_2$,...,$S^{'}_l$ of $Y^{'}$. Each subset $S^{'}_i$ contains $|S_i|+|Y|^2$ elements that are the tuples for query $Q$ from source $S_i$ and the newly generated ones.

For a query $Q$ of $|Y|+|Y|^2$ tuples, if there exist an optimal permutation $\Pi_{opt}(Y)$ of $m$ sources, we can easily see that the union of all elements in these sources cover all the elements of the universal set. Thus, the reduction from TMP to Set Cover Problem is established. In turn, suppose that there exist smallest number of $m$ subsets $S^{'}_{c(1)}$, $S^{'}_{c(2)}$, ..., $S^{'}_{c(m)}$ whose union equals the universal set $Y^{'}$. We firstly construct a permutation $\Pi(Y)$ by randomly arranging these $m$ subsets, and set $ta_i=0 \; ms$ and $tr_i=1 \; ms$ for $\forall S_i \in Y$. Then, we have the time cost for query $Q$ of $|Y|+|Y|^2$ tuples :
\begin{equation}
T(Q_{|Y|+|Y|^2}(\Pi(Y))) \leq m(|Y|+|Y|^2)
\end{equation}
Consider any other permutation $\Pi^{'}(Y)$ with $m+1$ subsets :
\begin{equation}
T(Q_{|Y|+|Y|^2}(\Pi^{'}(Y))) \geq (m+1)|Y|^2
\end{equation}
For this instance of TMP, we have $T(Q_{|Y|+|Y|^2}(\Pi(Y))) \leq T(Q_{|Y|+|Y|^2}(\Pi^{'}(Y)))$. Hence, the optimal permutation $\Pi_{opt}(Y)$ with $m$ subsets exists if the union of these $m$ subsets covers the universal set.

The decision version of TMP is to decide the query time cost for a given permutation $\Pi(Y)$. The total time cost for $\Pi(Y)$ can be calculated out in $O(l)$ time. Therefore, TMP is NP-complete.
\end{IEEEproof}

Since TMP is NP-complete, the basic solution of traversing all permutation of sources will soon become unmanageable when the number of sources increase. We consider an alternative scalable algorithm to construct the permutation for TMP. Naturally, to incrementally construct the permutation $\Pi(Y)$, a greedy algorithm can be planned to sequentially choose a source $S_i$ that has the fastest query rate $v_i=\frac{|S_i|-|\cap S_i|}{ta_i+tr_i |S_i|}$ at each iteration. For the example of Figure \ref{f1}, source $S_1$ is first selected since its query rate $v_1=1.43$ $tuple/ms$ is the fastest among these three sources. Afterward, source $S_2$ is then extracted with $v_2=\frac{125-35}{125\times 1.1}=0.65$ $tuple/ms$. Finally, source $S_3$ is chosen with $v_3=\frac{75-5-10}{75\times 1.5}=0.53$ $tuple/ms$. Another approach is to examine the residual tuples of each source and choose the source with maximal residual tuples at each iteration. For the example of Figure \ref{f1}, source $S_2$ is first chosen since it has the maximal residual tuples $|S_2|=125$. Then, source $S_3$ and $S_1$ are selected sequentially with residual tuples of $65$ and $10$ respectively. However, as shown in Example \ref{e_example}, neither the permutation $S_1S_2S_3$ by greedily considering fastest query rate nor $S_2S_3S_1$ by greedily considering maximal residual tuples is an optimal solution.

Actually, we investigate both the effect of query rate and residual tuples on source permutation for TMP. As can be seen in Figure \ref{f2}, curve $S_1S_2S_3$ returns more tuples compared to curve $S_2S_3S_1$ with the same time before the crosspoint of $(96.8, 106.4)$, and the result reverses after the crosspoint. For a given query $Q_k$, we could find a better permutation by discovering possible crosspoints. Next, we give two observations that reveal the correlation of query rate and residual tuples on crosspoint appearance.

\begin{figure}[!t] \centering
\subfigure[$S_1$ has faster query rate while $S_2$ has more residual tuples] { \label{fig4:a}
\raggedleft
\includegraphics[width=2.58in,height=1.8in]{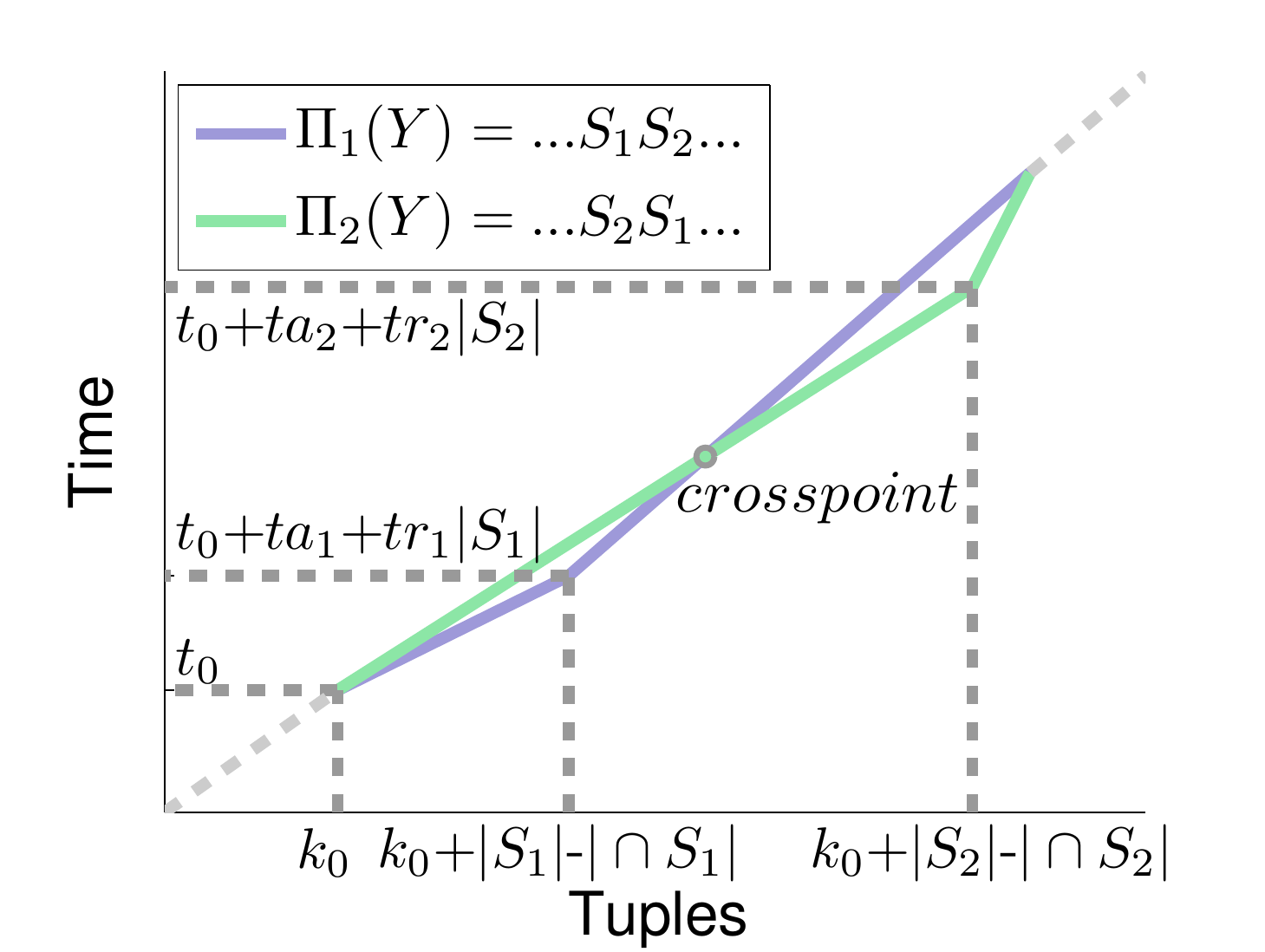}
}
\subfigure[$S_1$ and $S_2$ have few intersection tuples] { \label{fig4:b}
\raggedleft
\includegraphics[width=2.58in,height=1.8in]{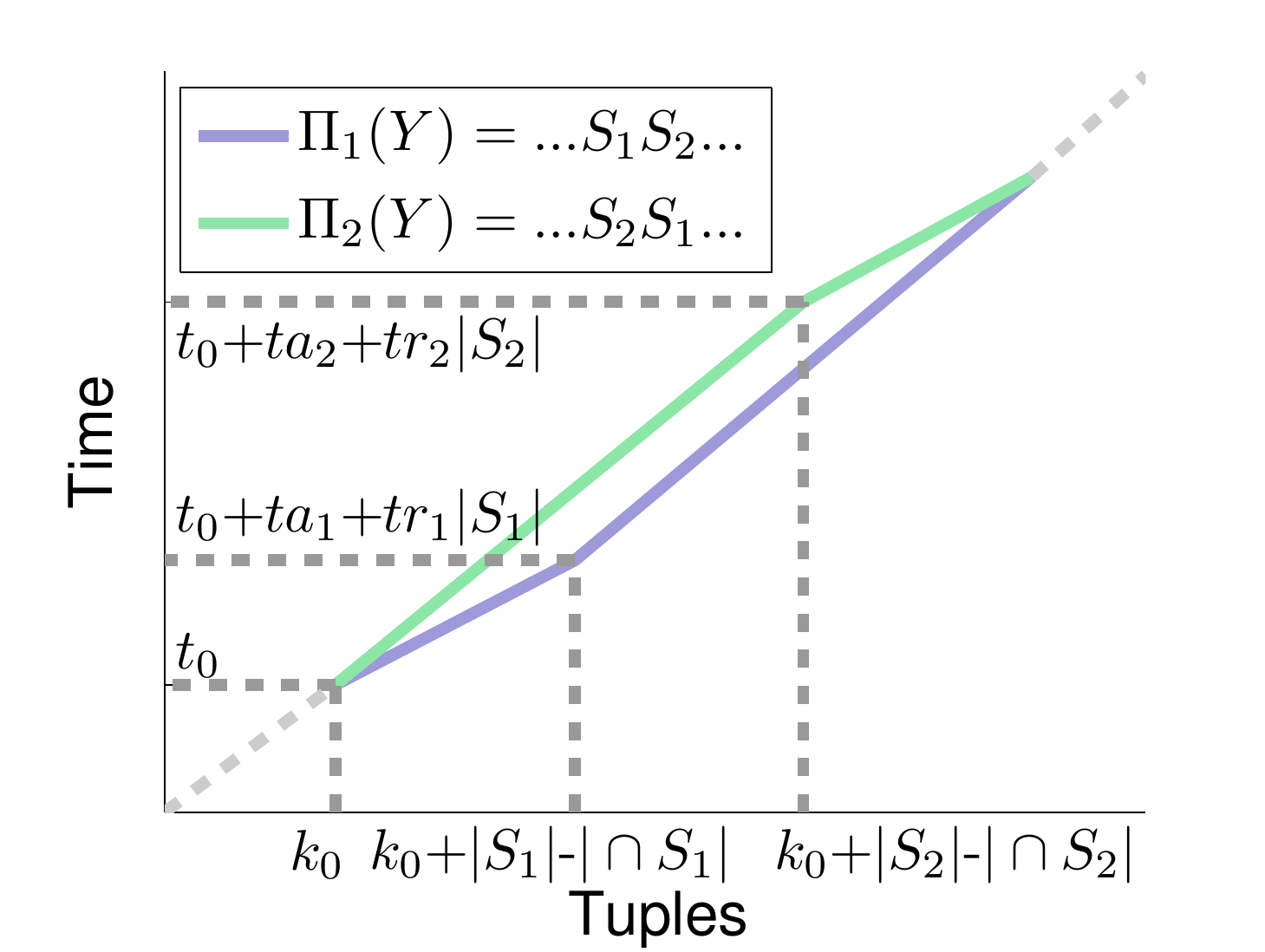}
}
\caption{Example of Observation \ref{obsv1} and Observation \ref{obsv2}}
\label{fig4}
\end{figure}

\begin{obsv} \label{obsv1}
The crosspoint of two permutation curves appears when a swap happens between a source with faster query rate and another source with more residual tuples.
\end{obsv}
In Figure \ref{fig4:a}, source $S_1$ has faster query rate and less residual tuples compared to source $S_2$ at time $t_0$. The query rate of the curve $\Pi_1(Y)$ slows down after $S_1$ has been queried while the query rate of the curve $\Pi_2(Y)$ keeps and catches up $\Pi_1(Y)$ at that time since $S_2$ has more tuples than $S_1$. The crosspoint of curve $\Pi_1(Y)$ and $\Pi_2(Y)$ exists when $S_2$ has enough more tuples and a little less query rate than $S_1$.

\begin{obsv} \label{obsv2}
The crosspoint of two permutaion curves does not exist when a swap happens between two sources that have few intersection tuples.
\end{obsv}
For the example in Figure \ref{fig4:b}, the query rate $v_2$ of $S_2$ has $v_2(\Pi_2(Y))-v_2(\Pi_1(Y))=\frac{|S_1 \cap S_2|}{ta_2+tr_2 |S_2|}$ for the curve $\Pi_1(Y)$ and $\Pi_2(Y)$. If source $S_1$ has fewer intersection tuples with $S_2$, that is, $|S_1 \cap S_2|$ has a small value, we have $v_2(\Pi_1(Y))\approx v_2(\Pi_2(Y))$ and $v_1(\Pi_1(Y))\approx v_1(\Pi_2(Y))$ for $S_1$. In this case, the crosspoint of these two curves does not exist.


The Observation \ref{obsv1} and \ref{obsv2} indicate heuristics for designing the Online Permutation (OnlinePerm) Algorithm. Before describing it, we firstly present the inputs, functions and algorithms on which the OnlinePerm algorithm is built. Given a query $Q$ of $k$ tuples, the following inputs provide essential information for the construction of source permutation:
\begin{itemize}
  \item Statistics Input $SI$: The statistics are from the set $Y=\{S_1,S_2,...,S_l\}$ of sources, including the access time cost $ta_i$, time cost $tr_i$ of querying a tuple, the number of tuples $|S_i|$ answering the given query $Q$ and the number of intersection tuples $|\cap S_i|$ for any permutation of sources, $i=1,2,...,l$. (Let $|\cap Y|$ to denote $|S_i|$ and $|\cap S_i|$ for $S_i \in Y$. $|\cap Y|$ is unknown or imprecise upon receiving $Q$ at runtime. In Section \ref{sc}, we will show how to estimate $|\cap Y|$ by two-stage detection);
  \item Intersection Threshold $\theta_{sp}$: For a source $S_i$, its intersection proportion with another source $S_j$ can be computed by $\frac{|S_i \cap S_j|}{|S_i|}$. If the value of $\frac{|S_i \cap S_j|}{|S_i|}$ is below the threshold $\theta_{sp}$, we can ignore the source $S_j$ for $S_i$ in the permutation construction.
\end{itemize}

\begin{algorithm} \label{alg_greedyQR}
\caption{Greedy on Query Rate (GreedyQR)}
\KwIn{A query $Q_k$; A permutation $\Pi(Y)$ of selected sources; A set $Y_e$ of unselected sources; Statistics Input $SI$}
\KwOut{The new permutation $\Pi(Y)$ and set $Y_e$; The total number of tuples $n_{sum}$ of $\Pi(Y)$; The average query rate $v_{avg}$ of $\Pi(Y)$}
$[n_{sum}]=Counter(\Pi(Y), SI)$;\\
\eIf{$n_{sum} \geq k$}{
    $[n_{sum},v_{avg}]=Perm2Set(\Pi(Y),Y_e,SI,k)$;\\
}{
    Set a empty source $S_q$ with no tuple;\\
    \While{$n_{sum} < k$}{
        Set $v_{max}=0$; \\
        $[\cap S_i, S_i \in Y_e]=InSec(\Pi(Y), Y_e, SI)$;\\
        \ForEach{Data Source $S_i \in Y_e$}{
            Calculate the query rate $v_i=\frac{|S_i|-|\cap S_i|}{ta_i+tr_i|S_i|}$;\\
            \If{$v_i>v_{max}$}{
                Set $v_{max}=v_i$ and $S_q=S_i$; \\
            }
        }
        $[n_{sum},v_{avg}]=Set2Perm(S_q,\Pi(Y),Y_e,SI)$;\\
    }
}
\Return $\Pi(Y)$, $Y_e$, $n_{sum}$ and $v_{avg}$;
\end{algorithm}

We define the following functions:
\begin{itemize}
  \item $InSec$: Given a permutation $\Pi(Y)$ of selected sources for the query $Q_k$ and a set $Y_e$ of unselected sources, function $InSec$ returns the number of intersection tuples $|\cap S_i|$ with the sources in $\Pi(Y)$ for $S_i \in Y_e$.
  \item $Counter$: Given a permutation $\Pi(Y)$ of selected sources for the query $Q_k$, function $Counter$ counts the total number of tuples $n_{sum} = \sum_{S_i \in \Pi(Y)} (|S_i|-|\cap S_i|)$;
  \item $Perm2Set$: Supposing that $ \sum_{i=1}^{i_k-1} (|S_i|-|\cap S_i|) < k \leq \sum_{i=1}^{i_k} (|S_i|-|\cap S_i|)$, the function $Perm2Set$ removes the $(i_k+1)$th, $(i_k+2)$th, ..., $|\Pi(Y)|$th sources from $\Pi(Y)$, and add these sources to the set $Y_e$ of unselected ones. Then, $Perm2Set$ returns the total number of tuples $n_{sum}$ and the average query rate $v_{avg}$ of $\Pi(Y)$;
  \item $Set2Perm$: Given the permutation $\Pi(Y)$ of selected sources and a set $Y_e$ of unselected sources, the function $Set2Perm$ adds a source $S_q \in Y_e$ to the tail of the $\Pi(Y)$, removes $S_q$ from $Y_e$, and returns the total number of tuples $n_{sum}$ and the average query rate $v_{avg}$ of $\Pi(Y)$;
  \item $Swap$: Given $S_i \in \Pi(Y)$ and $S_j \in Y_e$, the function $Swap$ replaces $S_i$ with $S_j$ in $\Pi(Y)$, removes the sources ranked behind $S_i$ from $\Pi(Y)$, and adds $S_i$ and these removed sources to $Y_e$. Then, $Swap$ returns the new $\Pi(Y)$ and $Y_e$;
  \item $Sort$: Given $S_i \in \Pi(Y)$ and an expression $Exp$, the function $Sort$ sorts the sources in $Y_e$ by the value of $Exp$, and discard the sources with the value below the threshold $\theta_{sp}$. Then, $Sort$ returns the sorted permutation $\Pi_s(Y_e)$ for $Y_e$;
  \item $LockWrite$: The function $LockWrite$ writes $\Pi(Y)$ to the shared memory for the query $Q_k$ with the update lock;
\end{itemize}

The Algorithm \ref{alg_greedyQR}, Greedy on Query Rate (GreedyQR), invokes the function $Counter$ to check the total number of tuples $n_{sum}$ in $\Pi(Y)$. If $n_{sum}\geq k$, GreedyQR moves extra sources from $\Pi(Y)$ to $Y_e$ by $Perm2Set$; else greedily selects the source with the maximal query rate and add it to the tail of $\Pi(Y)$ by $Set2Perm$ until $n_{sum}\geq k$.

\begin{algorithm} \label{alg_rePerm}
\caption{Re-Permutations (RePerm)}
\KwIn{A selected source $S_i$; A query $Q_k$; A permutation $\Pi(Y)$ of selected sources; A set $Y_e$ of unselected sources; The average query rate $v_{avg}$ of $\Pi(Y)$; Statistics Input $SI$; Intersection Threshold $\theta_{sp}$}
\KwOut{The rebuilt permutation $\Pi^{'}(Y)$; Set $Y^{'}_e$; Query rate $v_{avg}^{'}$}
Set $v_{avg}^{'}=0$; \\
$[\Pi_s(Y_e)]=Sort(S_i,Y_e,\frac{|S_i\cap S_j|}{|S_i|},\theta_{sp},"descending")$, $S_i \in \Pi(Y)$ and $S_j \in Y_e$;\\
\For{$j=1,2,...,|\Pi_s(Y_e)|$}{
    Fetch the $j$th source $S_j$ from $\Pi_s(Y_e)$;\\
    \If{$|S_i|<|S_j|$}{
        $[\Pi^{''}(Y),Y^{''}_e)]=Swap(S_i,S_j,\Pi(Y),Y_e)$;\\
        $[\Pi^{''}(Y),Y^{''}_e,n_{sum}^{''},v_{avg}^{''}]=GreedyQR(Q,k,\Pi^{''}(Y),Y^{''}_e,SI)$;\\
        \If{$v_{avg}^{''}>v_{avg}^{'}$}{
            Set $\Pi^{'}(Y)=\Pi^{''}(Y)$, $Y^{'}_e=Y^{''}_e$ and $v_{avg}^{'}=v_{avg}^{''}$;\\
        }
    }
}
\Return $\Pi^{'}(Y)$, $Y^{'}_e$ and $v_{avg}^{'}$;
\end{algorithm}

The Algorithm \ref{alg_rePerm}, Re-Permutations (RePerm), explores the possible swaps between a selected source $S_i \in \Pi(Y)$ and these sources in $Y_e$. Based on the Observation \ref{obsv2}, RePerm firstly selects the sources from $Y_e$ by sorting with the value of $\frac{|S_i\cap S_j|}{|S_i|}$ in descending order, and ignores the source $S_j$ if $\frac{|S_i\cap S_j|}{|S_i|}<\theta_{sp}$, $\forall S_j \in Y_e$. Then, RePerm only chooses the source $S_j$ having $|S_i|<|S_j|$ to swap $S_i$ by the knowledge from Observation \ref{obsv1}. After the swap process, RePerm re-ranks $S_i$, the sources that are originally ranked behind $S_i$ in $\Pi(Y)$, and the sources in $Y_e$ by GreedyQR to construct a candidate permutation $\Pi^{''}(Y)$. Finally, RePerm returns the permutation $\Pi^{'}(Y)$ with the maximal average query rate $v_{avg}^{'}$ from all candidate permutations.

The OnlinePerm algorithm (Algorithm \ref{alg_onlinePerm}) is proposed based on the effect of both the query rate and the residual tuples. If the given permutation $\Pi(Y)$ is empty, the OnlinePerm algorithm firstly construct the permutation $\Pi(Y)$ by GreedyQR that greedily adds the source with the maximal query rate from $Y_e$ sequentially, and invokes $LockWrite$ to write $\Pi(Y)$ to the shared memory for query $Q$ (The process of the Query Execution component that works in parallel with the OnlinePerm algorithm uses $LockRead$ to read the permutation $\Pi(Y)$, and queries on sources sequentially following $\Pi(Y)$). Since the sources that are more closely to the head of $\Pi(Y)$ are queried firstly, OnlinePerm checks and swaps the sources in $\Pi(Y)$ sequentially by RePerm from its head to tail. As presented above, RePerm tries to find a new permutation $\Pi^{'}(Y)$ based on the knowledge of the correlation between query rate and residual tuples provided by Observation \ref{obsv1} and \ref{obsv2}. Once a new permutation $\Pi^{'}(Y)$ having the average query rate $v_{avg}^{'}>v_{avg}$ is found, OnlinePerm invokes $LockWrite$ to write $\Pi^{'}(Y)$ to the shared memory.

Construct a permutation $\Pi_u(Y)$ that arranges sources by query rate $v_{u(i)}=\frac{|S_{u(i)}|}{ta_{u(i)}+tr_{u(i)}|S_{u(i)}|}$ in descending order (without considering the intersection tuples, e.g., $|\cap S_{u(i)}|$), where $S_{u(i)}$ denotes the $i$th source in $\Pi_u(Y)$. Let $u(i_k)$ denote the number of sources that has $\sum_{i=1}^{u(i_k)}|S_{u(i)}| \geq k$ and $\sum_{i=1}^{u(i_k)-1}|S_{u(i)}| < k$. We have the following theorem for $Q_k$:
\begin{thm} \label{t_approximation}
The time cost $T(Q_k(\Pi(Y)))$ of $\Pi(Y)$ constructed by OnlinePerm is $\frac{k \sum_{i=1}^{l} (ta_{i}+tr_{i}|S_{i}|)}{|Y| \sum_{i=1}^{u(i_k)} (ta_{u(i)}+tr_{u(i)}|S_{u(i)}|)}$-approximate to the optimal time cost $T(Q_k(\Pi_{opt}(Y)))$ for TMP.
\end{thm}
\begin{IEEEproof}
The time cost $T(Q_k(\Pi_{opt}(Y)))$ is greater than $T(Q_k(\Pi_u(Y)))$ that ignores the intersection tuples :
\begin{equation} \label{e_a1}
T(Q_k(\Pi_{opt}(Y))) \geq \sum_{i=1}^{u(i_k)} (ta_{u(i)}+tr_{u(i)}|S_{u(i)}|)
\end{equation}
GreedyQR has the monotonicity property that $v_i>v_j$ for sources $S_i$ ranked ahead of $S_j$ in permutation $\Pi(Y)$. OnlinePerm performs as well as GreedyQR at worst with no swap happening. Thus, according to the monotonicity property, we have :
\begin{equation} \label{e_a2}
\frac{k}{T(Q_k(\Pi(Y)))} \geq \frac{|Y|}{\sum_{i=1}^{l} (ta_{i}+tr_{i}|S_{i}|)}
\end{equation}
Combine (\ref{e_a1}) and (\ref{e_a2}) :
\begin{equation} \label{e_a3}
\frac{T(Q_k(\Pi(Y)))}{T(Q_k(\Pi_{opt}(Y)))} \leq \frac{k\sum_{i=1}^{l} (ta_{i}+tr_{i}|S_{i}|)}{|Y|\sum_{i=1}^{u(i_k)} (ta_{u(i)}+tr_{u(i)}|S_{u(i)}|)}
\end{equation}
\end{IEEEproof}

\begin{cor}
Given $ta_i \ll tr_{i}|S_i|$ and $tr_{i} \approx tr_{j}$ for $\forall S_i, S_j \in Y$, the time cost $T(Q_k(\Pi(Y)))$ of $\Pi(Y)$ by OnlinePerm is $\frac{\sum_{i=1}^{l} |S_i|}{|Y|}$-approximate to $T(Q_k(\Pi_{opt}(Y)))$ for TMP.
\end{cor}
\begin{IEEEproof}
With the given $ta_i \ll tr_{i}|S_i|$ and $tr_{i} \approx tr_{j}$, the rate of querying any number of tuples with the permutation $\Pi_u(Y)$ is approximately close to a constant :
\begin{equation} \label{e_a4}
\frac{k}{\sum_{i=1}^{u(i_k)} (ta_{u(i)}+tr_{u(i)}|S_{u(i)}|)} \approx \frac{1}{tr_{u(i)}}
\end{equation}
Combine (\ref{e_a3}) and (\ref{e_a4}) :
\begin{equation}
\frac{T(Q_k(\Pi(Y)))}{T(Q_k(\Pi_{opt}(Y)))} \leq \frac{\sum_{i=1}^{l} |S_i|}{|Y|}
\end{equation}
\end{IEEEproof}

\begin{algorithm} \label{alg_onlinePerm}
\caption{Online Permutations (OnlinePerm)}
\KwIn{A query $Q_k$; A permutation $\Pi(Y)$; Statistics Input $SI$; Intersection Threshold $\theta_{sp}$}
\KwOut{The new permutation $\Pi(Y)$ for query $Q_k$}
Set the permutation $\Pi(Y)$ as an empty queue;\\
Set the set $Y_e=Y$;\\
\If{$\Pi(Y)==\phi$}{
    $[\Pi(Y),Y_e,n_{sum},v_{avg}]=GreedyQR(Q,k,\Pi(Y),Y_e,SI)$;\\
}
Execute $LockWrite(\Pi(Y))$;\\
\For{$i=1,2,...,|\Pi(Y)|$}{
    Fetch the $i$th source $S_i$ from $\Pi(Y)$;\\
    $[Pi^{'}(Y),Y^{'}_e,v_{avg}^{'}]=RePerm(S_i,Q,k,\Pi(Y),Y_e,v_{avg},SI,\theta_{sp})$;\\
    \If{$v_{avg}^{'}>v_{avg}$}{
        Set $v_{avg}=v_{avg}^{'}$ and $\Pi(Y)=\Pi^{'}(Y)$;\\
        Execute $LockWrite(\Pi(Y))$;\\
    }
}
\Return $\Pi(Y)$;
\end{algorithm}

OnlinePerm sequentially checks the possible swaps between $\forall S_i \in \Pi(Y)$ and $S_j \in Y_e$ having $\frac{|S_i\cap S_j|}{|S_i|} \geq \theta_{sp}$ from the head to the tail of $\Pi(Y)$. Both the line 7-14 of Algorithm \ref{alg_onlinePerm} and the line 3-12 of Algorithm \ref{alg_rePerm} runs in $O(l)$ time. The time complexity of GreedyQR (Algorithm \ref{alg_greedyQR}) is $O(l^2)$. Therefore, OnlinePerm has time complexity $O(l^4)$. Given some restrictions on $\theta_{sp}$, e.g., set $\theta_{sp}(S_i)$ equal to the maximal value of $\frac{|S_i\cap S_j|}{|S_i|}$, $\forall S_j \in Y_e$, the time complexity of OnlinePerm can be reduced to $O(l^3)$.


\section{Statistics Collection} \label{sc}

The statistics inputs $SI$ described in the last section should be collected from sources to the querier so as to be taken as inputs for OnlinePerm. The SC component (figure \ref{f3}) can obtain the access time cost $ta_i$ and the per-tuple transfer time cost $tr_i$ of $SI$ for each source $S_i \in Y$ by generating queries on $S_i$. Then, $ta_i$ and $tr_i$ can be calculated by the time cost of retrieving result tuples, e.g., it takes $850 \; ms$ to retrieving $1000$ tuples from $S_i$, written as $ta_i + 1000tr_i = 850\; ms$. However, SC cannot precisely predict and pre-detect $|\cap Y|$ of $SI$ for a query $Q_k$ given in real time. Alternatively, SC solves this problem in two stages. In Initial-detection stage, SC generates a query $\widehat{Q}_{|\widehat{Y}|}$ that retrieves all tuples from all sources (or samples from all sources), and estimate $|\cap \widehat{Y}|$ for $\widehat{Q}_{|\widehat{Y}|}$. The query result of any other query can be regarded as a subset of the result of $\widehat{Q}_{|\widehat{Y}|}$. Thus, in the Online-detection stage, the estimation of $|\cap Y|$ for a real-time given query $Q_k$ can be derived both from $|\cap \widehat{Y}|$ and online detection results.

\subsection{Initial-detection}
In the Initial-detection stage, SC generates a query $\widehat{Q}_{|\widehat{Y}|}$ to retrieve all tuples from all sources in $Y$. For the sake of simplicity, we abuse notation and let $|\widehat{S_i}|$ and $|\widehat{S'_i}|$ denote the number of tuples in and not in $S_i$ for query $\widehat{Q}_{|\widehat{Y}|}$ respectively. Combine all $|\widehat{S_i}|$ or $|\widehat{S'_i}|$, $i=1,2,...,l$, as variables in set $\widehat{\Omega}$, e.g., $|\widehat{S}_1\widehat{S}'_2\widehat{S}'_3|$ (for $l=3$) as a variable denotes the number of tuples in $S_1$ and not in $S_2$, $S_3$. Also, Let $\widehat{\Omega}_q$ represent the set of variables that counts tuples in $q$ sources and not in the other $l-q$ sources. For a variable $\widehat{w}_{q+1} \in \widehat{\Omega}_{q+1}$, a variable $\widehat{w}_q$ is a parent of $\widehat{w}_{q+1}$ if $\widehat{w}_q$ and $\widehat{w}_{q+1}$ are different ("in" or "not in") in only one source, e.g., $|\widehat{S}_1\widehat{S}'_2\widehat{S}'_3| \in \widehat{\Omega}_1$ is a parent of $|\widehat{S}_1\widehat{S}_2\widehat{S}'_3| \in \widehat{\Omega}_2$. Let $|\widehat{S_i}|$ be the ancestor of all variables that consider tuples in $S_i$. As refer to $|\cap \widehat{Y}|$, given a permutation $\Pi(\widehat{Y})$ of sources, the number of intersection tuples $|\cap \widehat{S}_i|$ for a source $S_i \in Y$ can be estimated by adding up all the $|\widehat{S}_i|$ and $|\widehat{S}_j|$ variables that consider tuples both in $S_i$ and $S_j$ for any $S_j$ prior to $S_i$ in $\Pi(\widehat{Y})$.

To derive the estimation of $|\cap \widehat{Y}|$, SC needs to detect the value of all these variables in $\widehat{\Omega}$. The set $\widehat{\Omega}$ has a total of $2^l$ variables; the detection complexity of all variables in $\widehat{\Omega}$ grows exponentially as $l$ increases. It is impossible to detect all variables when hundreds or even thousands sources exist. Therefore, SC applies the pruning techniques that (1) iteratively add variables to the detection set, (2) iteratively remove variables from the detection set, and (3) estimate the value of variables by Maximum Entropy \cite{Jaynes1957}.

In detail, SC firstly detects the value of $|\widehat{S}_1|$, $|\widehat{S}_2|$, ..., $|\widehat{S}_l|$ from sources, and start the iterations of $l$ times. Let $\widehat{W}_i$ denote the detection set containing variables added in the $i$th iteration. In the $(i+1)$th iteration, SC firstly removes the variables whose value are blow a given threshold $\theta_{sc}$ from $\widehat{W}_i$, and queries on the sources for the value of the remaining variables in $\widehat{W}_i$ after the removal. Then, SC considers the variables in $\widehat{\Omega}_{q+1}$. If a variable $\widehat{w} \in \widehat{\Omega}_{q+1}$ has $\sum Parent(\widehat{w})> \theta_{sc}$, the value of $\widehat{w}$ may also greater than $\theta_{sc}$. SC add such variables like $\widehat{w}$ to the detection set $\widehat{W}_{i+1}$. Suppose that all variables in $\widehat{W}_{i+1}$ have equal weight, and thus according to the principle of Maximum Entropy, SC can estimate the value of these variables in $\widehat{W}_{i+1}$ by solving the following MaxEnt problem :

\begin{equation} \label{e_sc}
\begin{cases}
\displaystyle \text {max: } -\sum_{\widehat{w} \in \widehat{W}_{i+1}} \widehat{w}\log \widehat{w} \\
\displaystyle \text {s.t. \;} |\widehat{S}_i|=\sum_{\;} \widehat{w}, \; \widehat{w} \in Ancestor(|\widehat{S}_i|) \;\&\&\; \widehat{w} \in \widehat{W} \\
\displaystyle \text {\;\;\;\;\;} \widehat{W} = \cup\{\widehat{W}_{j},j=1,2,...,i+1\} \\
\end{cases}
\end{equation}

By solving (\ref{e_sc}) with Lagrange multipliers \cite{Skilling1984}, SC can get the expected value of all variables in $\widehat{W}_{i+1}$. When the loop is finished, SC can further estimate $|\cap \widehat{Y}|$ for $\widehat{Q}_{|\widehat{Y}|}$ with the value of all variables in $\widehat{W}$, $\widehat{W} = \cup\{\widehat{W}_{j},j=1,2,...,l\}$.

\begin{exam}
Consider a simple scene of three sources $S_1$, $S_2$ and $S_3$, and assume that all variables to be determined have value greater than $\theta_{sc}$. SC firstly queries on these three sources for the value of $|\widehat{S}_1|$, $|\widehat{S}_2|$ and $|\widehat{S}_3|$. In the $1$st iteration, SC solves the MaxEnt problem for $\widehat{W}_{1}$ :
\begin{equation} \label{e_sc1}
\begin{cases}
\displaystyle \text {max: } -\sum_{j=1,2,3} \widehat{w}_j\log \widehat{w}_j \\
\displaystyle \text {s.t. \;} |\widehat{S}_1|=\widehat{w}_1, \;\;  |\widehat{S}_2|=\widehat{w}_2 \\
\displaystyle \text {\;\;\;\;\;\;\;} |\widehat{S}_3|=\widehat{w}_3 \\
\end{cases}
\end{equation}
Where we have $\widehat{w}_1=|\widehat{S}_1\widehat{S}'_2\widehat{S}'_3|$, $\widehat{w}_2=|\widehat{S}'_1\widehat{S}_2\widehat{S}'_3|$ and $\widehat{w}_3=|\widehat{S}'_1\widehat{S}'_2\widehat{S}_3|$. By solving (\ref{e_sc1}), SC can get the estimation of $\widehat{w}_1$, $\widehat{w}_2$ and $\widehat{w}_3$. With the assumption that all variables have value greater than $\theta_{sc}$, no variable is removed from $\widehat{W}_{1}$. In the $2$nd iteration,  SC detects the value of variable $\widehat{w}_1$, $\widehat{w}_2$ and $\widehat{w}_3$, and solves the MaxEnt problem for $\widehat{W}_{2}$ :
\begin{equation} \label{e_sc2}
\begin{cases}
\displaystyle \text {max: } -\sum_{j=4,5,...,9} \widehat{w}_j\log \widehat{w}_j \\
\displaystyle \text {s.t. \;} |\widehat{S}_1|=\widehat{w}_1+\widehat{w}_4+\widehat{w}_5, \;\;  |\widehat{S}_2|=\widehat{w}_2+\widehat{w}_6+\widehat{w}_7 \\
\displaystyle \text {\;\;\;\;\;\;\;} |\widehat{S}_3|=\widehat{w}_3+\widehat{w}_8+\widehat{w}_9 \\
\end{cases}
\end{equation}
Where we have $\widehat{w}_4=|\widehat{S}_1\widehat{S}'_2\widehat{S}_3|$, $\widehat{w}_5=|\widehat{S}_1\widehat{S}_2\widehat{S}'_3|$, $\widehat{w}_6=|\widehat{S}'_1\widehat{S}_2\widehat{S}_3|$, $\widehat{w}_7=|\widehat{S}_1\widehat{S}_2\widehat{S}'_3|$, $\widehat{w}_8=|\widehat{S}'_1\widehat{S}_2\widehat{S}_3|$ and $\widehat{w}_9=|\widehat{S}_1\widehat{S}'_2\widehat{S}_3|$. In the $3$rd iteration, SC solves the MaxEnt problem for $\widehat{W}_{3}=\{|\widehat{S}_1\widehat{S}_2\widehat{S}_3|\}$. Given a permutation $S_1S_2S_3$, the number of intersection tuples for any source could be estimated, e.g. $|\cap \widehat{S}_3|=|\widehat{S}'_1\widehat{S}_2\widehat{S}_3| +|\widehat{S}_1\widehat{S}'_2\widehat{S}_3| +|\widehat{S}_1\widehat{S}_2\widehat{S}_3|$.
\end{exam}

The formal description of Initial-detection algorithm is omitted here due to space limit. Initial-detection algorithm runs loop of $l$ times, each time with $|\widehat{W}_i|$ variables and $l$ constraints for the MaxEnt problem, $i=1,2,...,l$.

\subsection{Online-detection}
After the Initial-detection stage, an estimation of $|\cap \widehat{Y}|$ for $\widehat{Q}_{|\widehat{Y}|}$ has already been established. Then, a permutation $\Pi(\widehat{Y})=S_{(1)}S_{(2)}...S_{(l)}$ of all sources can be constructed by OnlinePerm based on $|\cap \widehat{Y}|$. Upon receiving a real-time query $Q$ of $k$ tuples, SC start the Online-detection stage, and derives the estimation of $|\cap Y|$ from $|\cap \widehat{Y}|$ on-the-fly simultaneously with the query execution.

Online-detection stage can be divided into two sub-stages. In the first sub-stage, the number of tuples $|S_{(i)}|$, $i=1,2,...,l$ for $Q_k$ are detected sequentially following the permutation $\Pi(\widehat{Y})$. When receiving partial results $|S_{(1)}|$, $|S_{(2)}|$, ..., $|S_{(q)}|$, SC estimates the number of tuples for other sources :
\begin{equation} \label{e_sc3}
|S_{(i)}|=\frac{\widehat{S}_{(i)}}{q} \sum_{j=1}^{q} \frac{S_{(j)}}{\widehat{S}_{(j)}}, \;\; i=q+1,q+2,...,l
\end{equation}
Rewrite the MaxEnt problem as follow :
\begin{equation} \label{e_sc4}
\begin{cases}
\displaystyle \text {max: } -\sum_{w \in W} w\log w \\
\displaystyle \text {s.t. \;} |{S}_{(i)}|=\sum_{\;} w, \; w \in Ancestor(|{S}_{(i)}|) \;\&\&\; w \in W \\
\displaystyle \text {\;\;\;\;\;} W = \cup\{\widehat{W}_{j},j=1,2,...,l\} \\
\end{cases}
\end{equation}
The MaxEnt problem in the Online-detection stage considers all variables introduced in the Initial-detection stage. By solving (\ref{e_sc4}), SC can get the expected value of all variables in $W$, and thus estimate $|\cap Y|$ for any given permutation $\Pi(Y)$.

In the second sub-stage, all the results of $|S_{(1)}|$, $|S_{(2)}|$, ..., $|S_{(l)}|$ have been received. SC sorts all variables $w \in W$ by their expected value $|w-\widehat{w}|$ in descending order, where $\widehat{w}$ denotes the value of $w$ estimated in the Initial-detection stage. Then, SC sequentially detects the value of these variables in $W$ following this order. Upon receiving partial results of detection, SC resolves the MaxEnt problem of (\ref{e_sc4}) for the estimation of $|\cap Y|$. The Online-detection is terminated when the query execution of $Q_k$ is finished or all the results of detection have been received.

The formal description of Online-detection algorithm is omitted here due to space limit. Online-detection algorithm resolves the MaxEnt problem with at most $|W|$ times; in the $i$th time, estimate the expected value of no more than $|W|-i+1$ variables.

\section{Online Query Strategy} \label{strategy}
\begin{figure}
\centering
\includegraphics[width=3.6in]{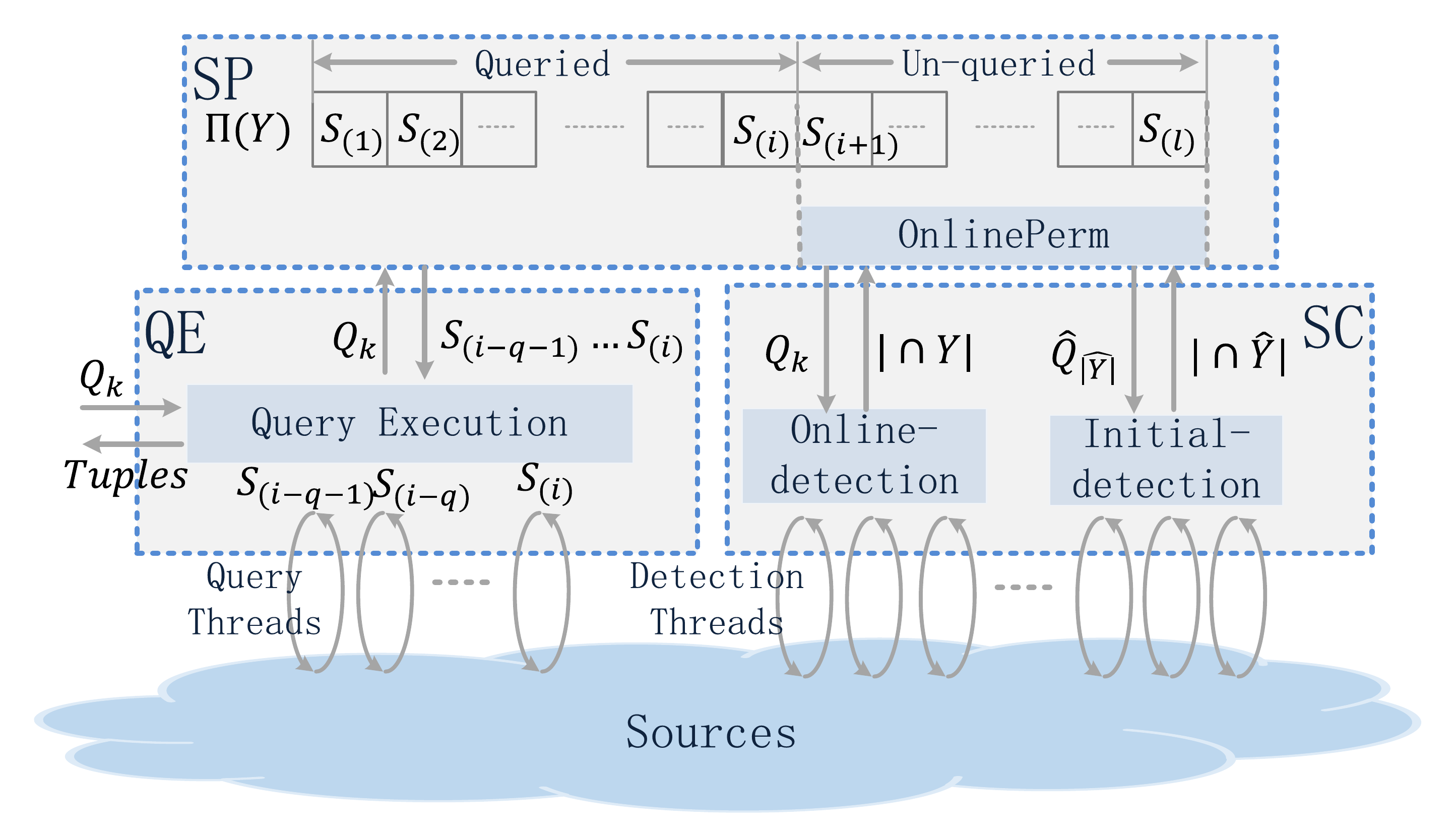}\\
\caption{\textrm{Online Query Framework}} \label{fig5}
\end{figure}

The Online Query Framework is shown in figure \ref{fig5}. It applies the dynamic strategy that enables the execution of query on sources and the improvement of source permutation to work in parallel.

At the beginning, SC generates a query $\widehat{Q}_{|\widehat{Y}|}$ that retrieves all tuples from all sources (or samples from all sources), and start the Initial-detection process to probe and estimate $|\cap \widehat{Y}|$. Then, SP applies OnlinePerm algorithm to construct a permutation $P(\widehat{Y})$ for all sources based on the estimation of $|\cap \widehat{Y}|$.

Upon receiving a real-time query $Q_k$, the processes in SC, SP and QE start to work simultaneously. In detail, SC starts the Online-detection process to derive the estimation of $|\cap Y|$ from both the results of online detection and the estimation of $|\cap \widehat{Y}|$. SP repeatedly runs the OnlinePerm algorithm on un-queried sources with the continuously new estimation of $|\cap Y|$ provided by SC; at the end of each run, the OnlinePerm algorithm writes the newly constructed permutation $\Pi(Y)$ to the shared memory for the query $Q_k$. Simultaneously, QE reads the permutation $\Pi(Y)$ from the shared memory, and starts query threads to retrieve the result tuples sequentially from sources following $\Pi(Y)$. When the querier has already received $k$ tuples or all sources have been queried, QE sends a signal to terminate the running process in SC and SP.

\section{Experiment Results} \label{experiment}

\begin{figure}
\centering
\includegraphics[width=2.4in]{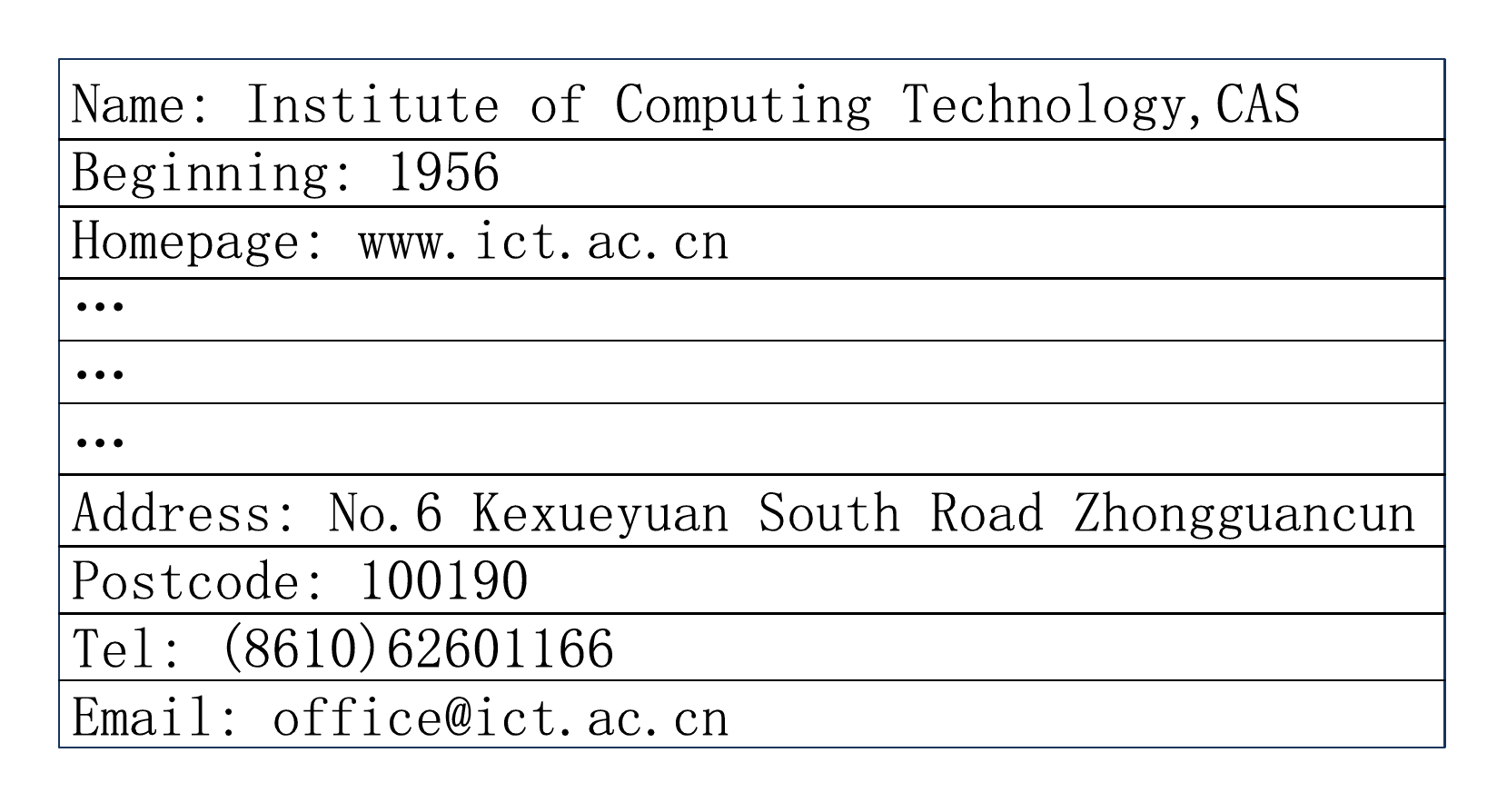}\\
\caption{\textrm{Example tuple}} \label{newfig_tuple}
\end{figure}

We conduct experiments on independent sources that are self-controlled and only provide query interface for others. Each source consists of tuples of educational institutions, as shown in Figure \ref{newfig_tuple}, crawled from web sites with a random start seed. In experiments, we divide tuples into two sets: $E_1$ and $E_2$, having $E=E_1 \cup E_2$ and $E_1 \cap E_2 = \emptyset$. Each tuple is either belong to $E_1$ or $E_2$. We use the query $\widehat{Q}$ of $SELECT \, * \, FROM \, E$ for the Initial-detection stage of SC and the query $Q$ of \texttt{$SELECT \; top \; k \; tuples \; FROM \; E_1$} to evaluate the time cost of implemented algorithms.

\begin{table*}[!t]
\renewcommand{\arraystretch}{1.3}
\caption{Experiment results} \label{experiment_results}
\centering
\begin{threeparttable}
\begin{tabular}{|c|p{1.3cm}|p{1.3cm}|p{1.3cm}|p{1.3cm}|p{1.3cm}|p{1.52cm}|p{1.52cm}|p{1.7cm}|}
  \hline
  \backslashbox{Condition}{Algorithm}  & Random & MaxT & MaxRT & MinT & MinRT & SeqPerm & OnlinePerm & FullKnowledge \\
  \hline
  $k=0.2|Y_1|$ & 38751.6\tnote{1} & 30478.7 & 26850.1 & 13807.9 & 11058.3 & 16617.4 & 10735.3 & 10528.6 \\
  \hline
  $k=0.4|Y_1|$ & 78872.4 & 65944.0 & 60727.7 & 32861.0 & 27899.4 & 31831.5 & 25709.8 & 24147.1 \\
  \hline
  $k=0.6|Y_1|$ & 133642.2 & 120663.7 & 109618.4 & 59887.5 & 52291.0 & 51415.3 & 47623.3 & 43979.9 \\
  \hline
  $k=0.8|Y_1|$ & 236555.6 & 209363.2 & 195653.2 & 114498.3 & 96686.0 & 90179.0 & 85064.9 & 81170.8 \\
  \hline
  $2$ query threads & 123385.5 & 111151.4 & 104554.5 & 60785.5 & 50789.2 & 52086.0 & 45509.7 & 42417.1 \\
  \hline
  $3$ query threads & 81871.1 & 74074.4 & 68009.5 & 38940.5 & 33843.2 & 35530.6 & 30339.8 & 28278.0 \\
  \hline
  $4$ query threads & 59615.4 & 54455.0 & 50029.3 & 30891.0 & 24287.4 & 27836.0 & 22754.8 & 21208.5 \\
  \hline
  $5$ query threads & 48772.5 & 41979.7 & 40106.5 & 24367.9 & 19719.2 & 23911.2 & 18203.8 & 17266.8 \\
  \hline
  $2000$ sources & 238312.2 & 208382.6 & 195475.3 & 114455.9 & 96382.1 & 91163.9 & 84995.8 & 81206.8 \\
  \hline
  $3000$ sources & 357818.7 & 300007.3 & 266163.6 & 169115.3 & 140972.3 & 131285.0 & 124255.4 & 118620.9 \\
  \hline
  $4000$ sources & 478313.0 & 393498.0 & 340143.2 & 221204.8 & 181479.6 & 167078.1 & 159867.4 & 152786.7 \\
  \hline
  $5000$ sources & 563392.5 & 481649.7 & 419245.2 & 273395.2 & 220551.2 & 202920.4 & 195332.8 & 186865.3 \\
  \hline
  $|Y_1|=0.2|Y|$ & 236438.5 & 232865.6 & 229671.7 & 149629.8 & 120992.3 & 103944.0 & 96441.3 & 81350.1 \\
  \hline
  $|Y_1|=0.4|Y|$ & 229253.9 & 217808.5 & 205961.6 & 133968.8 & 102889.0 & 95704.7 & 88965.9 & 80906.4 \\
  \hline
  $|Y_1|=0.6|Y|$ & 245076.4 & 202250.5 & 190735.5 & 107773.1 & 92073.5 & 89679.4 & 84411.5 & 81405.5 \\
  \hline
  $|Y_1|=0.8|Y|$ & 232241.0 & 191382.1 & 180240.4 & 103067.2 & 89814.8 & 87802.6 & 83070.8 & 80690.8 \\
  \hline
  $1.2times$ overhead & 237187.9 & 215087.3 & 201473.8 & 129091.8 & 112387.0 & 105885.1 & 93649.4 & 81543.8 \\
  \hline
  $1.4times$ overhead & 235221.0 & 223254.1 & 214012.1 & 145859.1 & 123970.6 & 123532.7 & 109257.6 & 82717.9 \\
  \hline
  $1.6times$ overhead & 235069.7 & 235806.8 & 227458.2 & 161720.6 & 132491.4 & 141180.2 & 124865.8 & 82069.3 \\
  \hline
  $1.8times$ overhead & 236087.7 & 235286.4 & 229887.0 & 164127.1 & 143741.9 & 158827.7 & 140474.1 & 81410.6 \\
  \hline
\end{tabular}
\begin{tablenotes}
    \footnotesize
    \item[1] All values are in milliseconds, or ms.
\end{tablenotes}
\end{threeparttable}
\end{table*}

For evaluation, we implemented eight algorithms:
\begin{itemize}
  \item Random: Randomly choose a permutation of sources.
  \item MaxT: Select the source $S$ with the maximal tuples $|S|$ each time without considering intersection between sources.
  \item MaxRT: Select the source $S$ with the maximal residual tuples $|S|-|\cap S|$ each time that considers the intersection compared with MaxT.
  \item MinT: Select the source $S$ with the minimal per-tuple retrieve time $\frac{ta+tr|S|}{|S|}$ each time without considering the intersection.
  \item MinRT: Select the source $S$ with the minimal residual per-tuple retrieve time $\frac{ta+tr|S|}{|S|-|\cap S|}$ each time that considers the intersection compared with MinT.
  \item SeqPerm: apply the sequential strategy that starts the query execution until the finish of RePerm (Algorithm \ref{alg_rePerm}). SeqPerm is with additional time cost of permutation construction, but may have a better permutation than OnlinePerm.
  \item OnlinePerm: apply the online strategy that let all the components work in parallel. OnlinePerm does not consider the sources have been queried, and only construct permutation for un-queried sources dynamically without waiting for the finish of RePerm.
  \item FullKnowledge: apply the permutation constructed by OnlinePerm with precise intersection statistics $|\cap Y|$ as input. Its performance can be considered as the upper bound can be achieved although the permutation may not be optimal (Theorem \ref{t_approximation}).
\end{itemize}

All the algorithms were implemented in Java JDK 1.7, and ran experiments on a Data Integration System (DIS) of our implementation. DIS can manipulate remote relational databases or shared folders to create a new source. DIS is built on $4$ physical machines. By default, DIS has $2035$ sources with a total of $501760$ tuples in these machines, and each tuple is limited with the maximal size of $120B$. Among these tuples, there are $24860$ distinct tuples in all. We observed that the access time cost of any source in DIS is in $[477, 2350]$ms, and the per-tuple transfer time cost is in $[0.02, 0.42]$ms. The querier was implemented on a Windows 7 machine with 2.3GHz Intel Core i7 CPU, 8GB RAM and Gigabit Ethernet Controller. Both the querier and DIS are in the same LAN. For simplicity, we abuse the notation and let $|Y_1|$, $|Y_2|$ and $|Y|$ denote the number of tuples in $E_1$, $E_2$ and $E$ respectively. By default, we set $|Y_1|=|Y_2|=0.5|Y|=12430$, $k=0.8|Y_1|$, the intersection threshold $\theta_{sp}=0.05$, the detection threshold $\theta_{sc}=0.005$, one thread running for the query execution and one thread running for the Online-detection process of $Q$ in the querier. We next evaluate the query time cost of implemented algorithms under conditions of different factors.

\subsection{Varying $k$}
We firstly compared the algorithms under the condition of varying $k$ by default settings. The row $2$-$5$ of Table \ref{experiment_results} show the results of query time cost of the algorithms for $k=0.2|Y_1|$, $0.4|Y_1|$, $0.6|Y_1|$ and $0.8|Y_1|$ respectively. We have the following observations.

First, Random performs worst among these algorithms. By considering the number of tuples in sources, MaxT has slightly less time cost than Random; By further considering the number of intersection tuples between sources, MaxRT performs better than MaxT. By selecting the sources with less per-tuple time cost, MinT is faster than MaxRT, and by considering the intersection in per-tuple time cost, MinRT has even faster query rate than MinT.

Second, SeqPerm has significantly initial time cost for permutation construction and performs worse than MinRT at the beginning. As shown in Table \ref{experiment_results}, SeqPerm has approximately $5.5$s (or seconds) and $3.9$s more time cost than MinRT for $k=0.2|Y_1|$ and $k=0.4|Y_1|$ respectively. Then, SeqPerm quickly catches up MinRT and performs better than MinRT; SeqPerm has approximately $0.9$s and $6.5$s less time cost than MinRT for $k=0.6|Y_1|$ and $k=0.8|Y_1|$ respectively. The results of SeqPerm show the effectiveness of RePerm algorithm that constructs a better permutation than MinRT.

Third, OnlinePerm has roughly the same performance with MinRT at the beginning, and as $k$ increases, OnlinePerm has apparently less time cost than MinRT. This is because that the effect of intersection between sources is little or non-existent when the number of tuples, $k$, to be retrieved is small, and the effect of intersection appears when $k$ is increased. As shown in Table \ref{experiment_results}, OnlinePerm and MinRT have approximately time cost of $10.7$s and $11.1$s for $k=0.2|Y_1|$ respectively, and OnlinePerm takes approximately $85.1$s to retrieve $k=0.8|Y_1|$ tuples while MinRT takes approximately $96.7$s.

Finally, as can be seen from Table \ref{experiment_results}, OnlinePerm has stable less time cost ($4$-$6$s) than SeqPerm for $k=0.2|Y_1|$, $0.4|Y_1|$, $0.6|Y_1|$ and $0.8|Y_1|$ respectively, which shows that the parallel execution of processes in SP and QE component would not reduce (or has little effect on) the quality of the permutation constructed by RePerm; OnlinePerm has little more time cost ($0$-$4$s) than FullKnowledge, which shows the efficiency of processes that online collect statistics for source permutation in SC component.

\subsection{Varying Query Threads}
We compared the algorithms on varying query threads. The row $6$-$9$ of Table \ref{experiment_results} show the results of query time costs of the algorithms for $k=0.8|Y_1|$ when running $2$, $3$, $4$ and $5$ query threads at the same time respectively. As can be seen, when more query threads are running, (1) the query time cost of the algorithms decrease fairly fast, (2) the decrease rate of query time cost slows down we usually observed in parallel system, and (3) the advantage of OnlinePerm on the performance becomes less apparent, e.g., OnlinePerm spends approximately $18.2$s on query execution than $19.7$s by MaxRT when $5$ query threads are running. The performance of OnlinePerm reveals that the benefit of Online-detection for statistics collection is banlanced by increased running query threads.

\subsection{Varying Sources}
We compared the algorithms on varying sources, and let the number of sources be $2000$, $3000$, $4000$ and $5000$ respectively by disabling or creating sources in DIS for evaluation. Then, we kept the number of distinct tuples and the total number of tuples as default settings by moving tuples between sources. The evaluation results of varying number of sources for $k=0.8|Y_1|$ are shown in row $10$-$13$ of Table \ref{experiment_results}. As can be seen, with more sources involved, (1) the query time of the algorithms increase, (2) the query time cost of OnlinePerm is less than other algorithms, and (3) the query time cost of OnlinePerm increases linearly and is approximate to the time cost of FullKnowledge, which shows the scalability of our online query system.

\subsection{The Effect of Pruning Techniques}
We applied the pruning techniques introduced in Section \ref{sc} for estimation of $|\cap Y_1|$ based on the statistics collected by the SC component. With huge number of variables removed by the pruning techniques, SC only need to solve the MaxEnt problem of $5137$ variables at the beginning of the Online-detection stage(, and the number of variable becomes less with the execution of Online-detection process). By converting the MaxEnt problem to a sparse system of linear equations, SC estimated the value of $5137$ variables in average $2.7$s as measured. In contrast, the basic approach without applying the pruning techniques should solve the MaxEnt of $2^{2035}$ variables. Obviously, this basic approach is uncomputable with so huge number of variables. Additionally, we measured the case of only $15$ sources. In this case, the basic approach spent $18.7$s to solve the MaxEnt while our approach with the pruning techniques finished in milliseconds.

\subsection{The Error of Initial-detection Statistics}
We compared the algorithms on errors of Initial-detection statistics. The statistics estimation for $Q$ is more accurate with a higher $\frac{|Y_1|}{|Y|}$, and we set $|Y_1|=0.2|Y|$, $0.4|Y|$, $0.6|Y|$ and $0.8|Y|$ respectively for evaluation. The evaluation results for $k=0.8|Y_1|$ are shown in row $14$-$17$ of Table \ref{experiment_results}. As can be observed, with higher $|Y_1|$, (1) the query time cost of the algorithms except random decrease, and especially (2) the decrease rate of query time cost of OnlinePerm is small, e.g., OnlinePerm spends approximately $84.4$s for $|Y_1|=0.6|Y|$ and $83.1$s for $|Y_1|=0.8|Y|$. The result of OnlinePerm shows that OnlinePerm has stable performance on errors of Initial-detection statistics.

\subsection{The Overhead of Online-detection}
We compared the algorithms on overheads of Online-detection by adding cycle time and setting the detection time cost to be $1.2$, $1.4$, $1.6$ and $1.8$ times of original detection time cost. The results are shown in row $18$-$21$ of Table \ref{experiment_results}. As can be observed from the results, with higher overhead of Online-detection, (1) the query time cost of the algorithms except random increase, and (2) OnlinePerm has a high increase rate of query time cost although it still performs better than other algorithms. The results reveal that it is getting harder to gain benefit from Online-detection when the detection overhead is increasing. In this case, more running threads for Online-detection are suggested to balance the effect of increased overhead.

\subsection{Summary}
We evaluated our online scheduling strategy under the condition of various factors.
\begin{itemize}
  \item Varying $k$: OnlinePerm is the fastest algorithm among all these algorithms; this evaluation result shows the efficiency of our strategy that enables SP, SC and QE working in parallel.
  \item Varying query threads: OnlinePerm still performs best among all these algorithms, but the benefit of Online-detection of SC is balanced by increased running threads.
  \item Varying sources: The query time cost of OnlinePerm increases linearly and least among all the algorithms when more sources are involved; this evaluation result shows that our strategy is scalable.
  \item The effect of pruning techniques: By applying the pruning techniques, SC can efficiently estimate the statistics of sources.
  \item The error of initial-detection statistics: OnlinePerm suffers a low performance degradation when the error of Initial-detection increases; this evaluation result shows that our strategy is efficient and robust.
  \item The overhead of online-detection: Although OnlinePerm still perform best among all these algorithms, it has a high performance degradation when detection overhead increases. This result suggests more detection threads of SC to balance the effect of increased detection overhead.
\end{itemize}

\section{Conclusion and Future Work} \label{conclusion}
We address the Time-Cost Minimization Problem (TMP) and propose the online scheduling strategy in this paper. The architecture of online query scheduling mainly contains three components of Source Permutation (SP), Statistics Collection (SC) and Query Execution (QE). We prove that it is NP-complete to construct a optimal permutation of sources and propose OnlinePerm algorithm that considers the effect of query rate and residual tuples for SP. We present a two-stage detection mechanism and apply pruning techniques to avoid the exponential number of variables estimation for SC. By applying the online scheduling strategy, SP, SC and QE work in parallel to reduce the total time cost for the query. The experiment results show the efficiency and scalability of our scheduling strategy.

In this paper, we simplify the redundant problem by only considering the repetitive data tuples between sources. In the future, we would concern the time cost of data fusion between partially overlapping tuples during online query scheduling.

\end{document}